\title{\bf Study in a beam test of the resolution of a Micromegas TPC
with standard readout pads}  
\author{D. C. Arogancia$^1$, A.M. Bacala$^1$, K. Boudjemline$^2$, D. R. Burke$^3$, \\
P. Colas$^3$, M. Dixit$^{2,13}$, K. Fujii$^4$, H. Fujishima$^{10}$, A. Giganon$^3$, I. Giomataris$^3$, \\
H. C. Gooc$^1$, M. Habu$^{7}$, T. Higashi$^{10}$, Y. Kato$^5$, M. Kobayashi$^4$, H. Kuroiwa$^4$, \\
V. Lepeltier$^6$, T. Matsuda$^4$, S. Matsushita$^{7}$, K. Nakamura$^{7}$, O. Nitoh$^7$,\\
 R. L. Reserva$^1$, Ph. Rosier$^8$, K. Sachs$^2$, R. Settles$^9$, A. Sugiyama$^{10}$, \\
 T. Takahashi$^{11}$, T. Watanabe$^{12}$, A. Yamaguchi$^{14}$, H. Yamaoka$^{4}$, Th. Zerguerras$^{8}$\\
${}^1$~\small{Department of Physics, MSU-Iligan Institute of Technology, Iligan City, Philippines}\\
${}^2$~\small{Carleton University, Ottawa, Canada}\\
${}^3$~\small{DAPNIA, CEA Saclay, 91191 Gif sur Yvette, France}\\
${}^4$~\small{IPNS, High Energy Accelerator Research Organization, Tsukuba, Japan}\\
${}^5$~\small{Kinki University, Japan}\\
${}^6$~\small{LAL, IN2P3-CNRS, et Universite de Paris-Sud 11, F91898, Orsay, France}\\
${}^7$~\small{Tokyo University of Agriculture and Technology, Japan}\\
${}^8$~\small{IPN, IN2P3-CNRS, et Universite Paris-Sud 11, F91406, Orsay, France}\\
${}^9$~\small{Max Planck Institute for Physics, Munich, Germany}\\
${}^{10}$~\small{Saga University, Japan}\\
${}^{11}$~\small{Hiroshima University, Japan}\\
${}^{12}$~\small{Kogakuin University, Japan}\\
${}^{13}$~\small{TRIUMF, Canada}\\
${}^{14}$~\small{University of Tsukuba, Japan}\\
}

\date{April 13, 2007}      

\documentclass[12pt]{articlecea}                                                   
\usepackage{epsfig}

\newcommand{\mathbold}[1]{\mbox{\boldmath $#1$}}

\newcommand{\gsim}{\hbox{ \raise3pt\hbox to 0pt{$>$}\raise-3pt\hbox{$\sim$} }}
\newcommand{\lsim}{\hbox{ \raise3pt\hbox to 0pt{$<$}\raise-3pt\hbox{$\sim$} }}
\newcommand{\del}{\ifmmode{\nabla}               \else{$\nabla$ }               \fi}

\newcommand{\figdir}{.}
					 
\begin{document}             
\maketitle                   
\begin{abstract}
The results of a beam test performed at the KEK PS in June 2005 are presented. 
Drift properties of an argon-isobutane mixture have been accurately measured and compared 
with predictions at magnetic fields between 0 and 1 Tesla. The r.m.s. point resolution of a 
padrow is compared with simulation and with an analytical calculation.
The fundamental limitations due to detector geometry and gas properties are reviewed
and the measured performances of the detector are found to be close to this limit.
A numerical application to the case of a Linear Collider TPC is presented.
\end{abstract}

\section{Introduction}      

Three of the four detector concepts which have been proposed for the Linear Collider foresee a large 
Time Projection Chamber (TPC)
as a main tracker. This allows continuous tracking to be performed, yet with a minimal amount of matter.
Depending on the detailed designs, the TPC should have about 200 padrows with a space resolution between 
100 and 150 microns in the R$\phi$ direction.
Mainly three technologies are currently considered for the gas amplification in these TPCs : a Multi-Wire 
Proportionnal Chamber (MWPC), a Micromegas chamber, and a multiGEM structure.
To this end, R\&D has been pursued since the beginning of the decade within the LC-TPC collaboration\cite{lctpc}
and has lead to the construction of several prototypes. It was felt useful to gather around a single experiment 
and to have a way of comparing the various technologies in a well-defined framework, with the same readout 
electronics and chamber geometry. This is the purpose of the Asian-Canadian-European Multi-Prototype 
collaboration. It started with the
construction of a MWPC chamber in MPI Munich, which took beam data at KEK in April 2004. Then the chamber  
was equipped with a triple-GEM structure and took beam data in April 2005. This paper reports on a beam test 
carried out with a Micromegas endplate in June 2005.

The detector and the operation conditions are described in Section 2, where an assessment of data quality is 
given. The results on gas properties (drift velocity, diffusion) are presented in Section 3.
Section 4 is devoted to the spatial resolution of the device.
It starts with a theory elaborated in our group which is then 
compared to simulations and measurements.
Fundamental limitations from the gas mixture and pad geometry are assessed and 
consequences for a Linear Collider TPC are drawn.

\section{Experimental Setup and Data Taking}

This experiment was carried out using the $\pi_2$ beam line at the KEK 12 GeV PS.
The beam line provided a secondary beam of pions or protons with
momenta up to 4~GeV/c through
the interaction of 12 GeV protons on a Be target, 
followed by the charge and momentum selection with a set of dipole magnets.
The beam spill has a flat top of $1.5$~s with a repetition rate of $0.25$~Hz.
There were 4 scintillation counters, TC1 through TC4, whose 4-fold coincidence triggered 
the data acquisition on 4~GeV negatively charged pions.
A typical trigger rate was $12$~Hz.
The first two trigger counters, TC1 and TC2, were placed at the entrance of the beam 
just downstream of a beam slit to control the beam intensity 
and had an overlap region of $2 \times 2$~cm$^2$.
The other two trigger counters, TC3 and TC4, were located at 8~m downstream of TC1 and TC2,
that is, just in front of our Multi-Prototype TPC (MP-TPC).
TC3 and TC4 had an overlap region of $30 \times 10$~cm$^2$ 
that matched the drift region of the chamber.
The MP-TPC together with TC3 and TC4\footnote{
TC3 and TC4 were equipped with fine-mesh photo multipliers, Hamamatsu R6682, 
which allowed operation in a magnetic field up to 1~Tesla.
} 
were installed in a 
Persistent Current solenoidal Magnet (PCMAG) having a bore diameter of
$85$~cm and a length of $1.3$~m with a very thin wall
of $20$~\% radiation lengths.
The magnet capable of creating a field up to 1.2 Tesla 
was operated in the closed loop mode and
provided a field uniformity better than $0.5$~\% in the drift region of the MP-TPC
that was aligned with the magnet axis, 
so that the electric drift field was parallel to the magnetic field.

In the following, unless otherwise stated, the beam was shot perpendicular
to the drift axis of the MP-TPC. 
The nominal size of the beam in these conditions is about 4 cm at the chamber.
To cover the whole active volume of the MP-TPC, 
we hence spread the beam by inserting a 5~cm-thick lead brick just downstream
of TC1 and TC2 in the normal data taking conditions.


\subsection{Multi-prototype TPC}
\begin{figure}                                                    
\center                                                          
\epsfig{figure=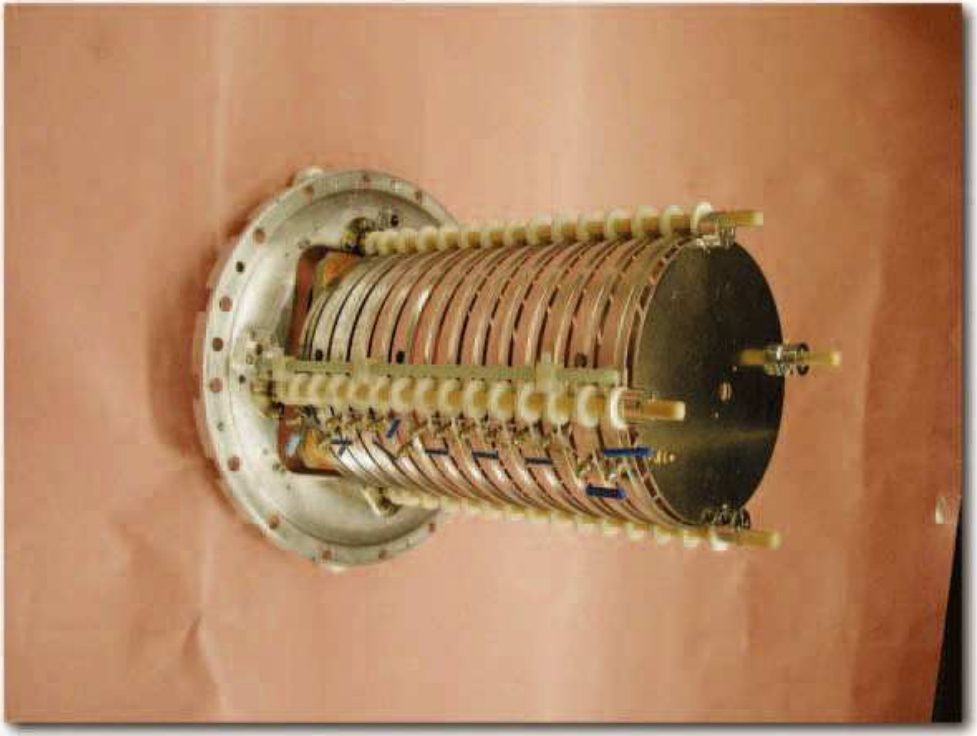,height=10cm} 
\caption{The Multi-Prototype TPC (MP-TPC)}   
\label{Fig:mptpc}
\end{figure}

The MP-TPC has 
a cylindrical drift region of 
261~mm
in length and 145 mm in diameter.
As seen in Fig.~\ref{Fig:mptpc}, 
the field cage is made of fourteen 
15~mm-wide circular rings spaced 3.2 mm apart and
a cathode disk with a small hole at its center\footnote{
The hole lets X-rays from an ${}^{55}$Fe 
source enter the drift volume, for gain calibration and monitoring purposes.
}.
All of these are embedded into a 
gas-tight vessel, which is closed on one side by
a detachable endplate detector that carries a Micromegas foil and readout pads.
The ring closest to the Micromegas detector is at a distance of 6~mm from the mesh.

The Micromegas endplate, built in Saclay and Orsay, 
consists of a 0.8 mm thick Printed Circuit Board bearing 
anode pads, with a mesh stretched on a frame and mounted above the pad plane.
The mesh is $5 \, \mu$m thick Cu having circular holes of $35 \, \mu$m in diameter
placed with a $60 \, \mu$m pitch. 
The $100 \times 100$ mm$^2$ frame leaves a $75 \times 75 $ mm$^2$ active area. 
A 50-micron gap is maintained between the mesh and
the pad plane by kapton pillars.
In normal operation a voltage of about 350~V is applied
to the mesh relative to the pad plane at 0~V.
The resulting electrostatic force sticks the mesh to the anode plane.
There are 12 rows of 32 pads on the anode plane.
Each pad has a rectangular shape and is placed
at a pitch of 6.3 mm along the beam ($y$ direction) 
and 2.3 mm transverse to the beam ($x$ direction). 

The chamber is filled with an argon
mixture containing 5\% isobutane and is operated at room temperature
and at atmospheric pressure.

\subsection{Readout Electronics and DAQ}
The 12 rows of 32 readout pads (384 in total) are
connected via $30$~cm long flat cables to 24 ALEPH preamplifiers\cite{Ref:ALEPH},
each having 16 channels and reading out a half of a pad row.
The shaped signals from the preamplifiers are sent to ALEPH TPDs (TPC Digitizers)
in a Fastbus crate via $15$~m long twisted pair cables and  are
sampled at a rate of 12.5 MHz and digitized over 8 bits.
The data are then read out, via a Fastbus-VSB translator unit, $FVSBI 9210$, 
by a VME board computer, $FIC 8234$, operating on OS/2.
The readout data are stored via TCP/IP connection on a Linux PC 
in the LCIO format\cite{Ref:LCIO}.
The data acquisition time is about $4\,$s per event, which limited the data
acquisition speed to about $1$ event per spill.

\subsection{Data Taking}
\begin{figure}                                                    
\center                                                          
\epsfig{figure=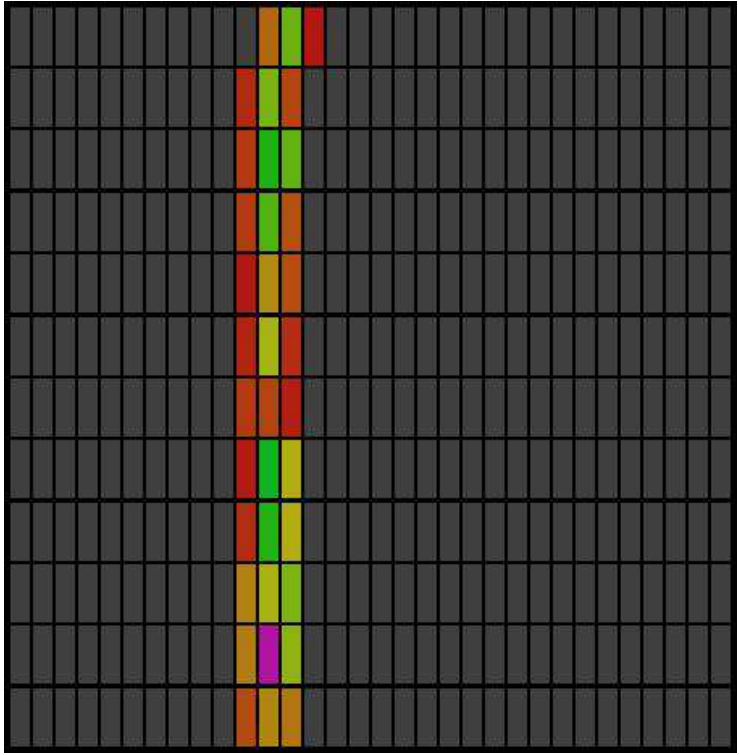,height=10cm} 
\caption{A typical 4 GeV pion track from the KEK $\pi_2$ beam seen by the Micromegas MP-TPC in a 0.5 T magnetic 
field ; the squares represent the pads hit with a colour code corresponding to the charge deposited.}  
\label{fig:evtdisp}
\end{figure}

A typical event taken at a magnetic field of 0.5 T is shown in Fig.~\ref{fig:evtdisp}.

\begin{figure}                                                    
\center                                                          
\epsfig{figure=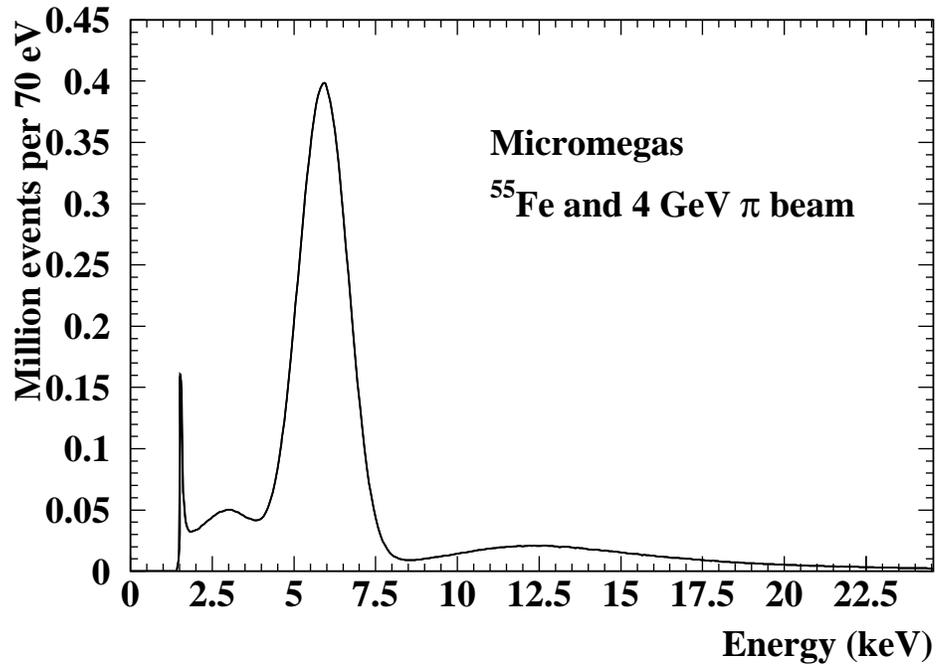,height=10cm} 
\caption{Energy distribution from the mesh signals from the source and from the beam}   
\label{fig:spect}
\end{figure}

\begin{figure}                                                    
\center                                                          
\epsfig{figure=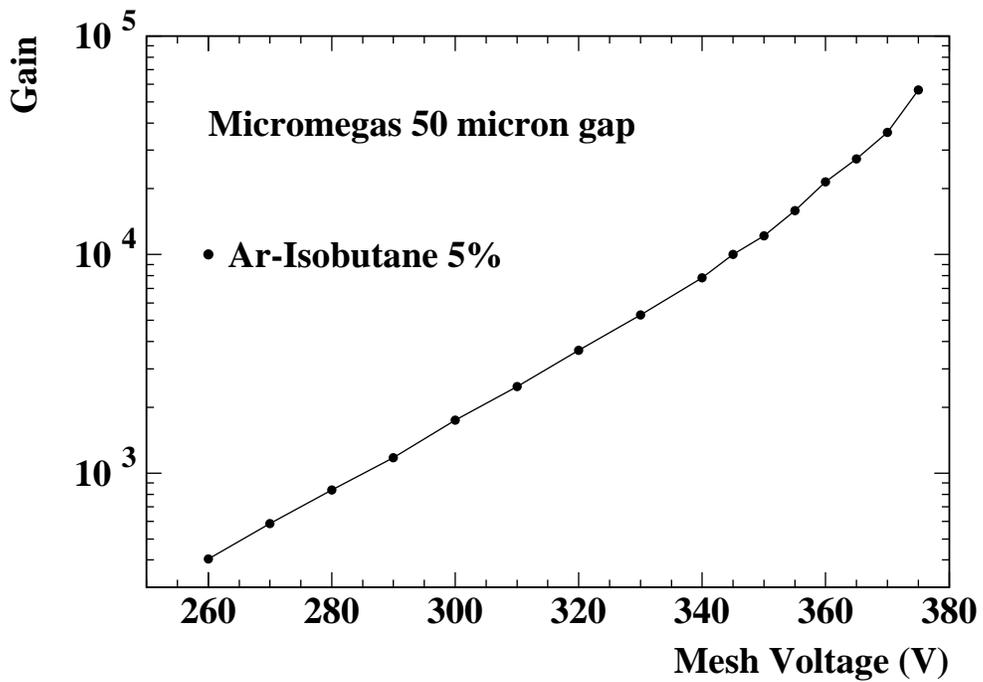,height=10cm} 
\caption{Gas gain as a function of the mesh voltage}   
\label{fig:gain}
\end{figure}

During the data taking, the gain was continuously monitored by a ${}^{55}$Fe source. 
The mesh signal was readout by a fast charge amplifier ORTEC 142B 
and the signal was sampled with an AMPTEK Multi-Channel Analyser MCA8000.
A spectrum obtained this way is shown in Fig.~\ref{fig:spect} 
where the 5.9 keV line and the escape line in argon are seen. 
The source was not collimated. 
The Landau distribution of the ionisation produced by the 4 GeV pions from the beam
is also visible and peaks at 12.5 keV as expected for the 75 mm active length of gas.
Over a period of 60 hours the extreme variations of gain were $\pm$3\% and the r.m.s. 
gain variations were 3 per mil over this period.
The gain is measured as a function of the mesh voltage (Fig.~\ref{fig:gain}). 
The data presented here  were taken at a mesh voltage of 320 V, 
corresponding to a gain of 3650, in a magnetic field of 0.5 and $1\,$T, and
a mesh voltage of 340~V (gain of 7500) for the data taken with no magnetic field.

A total of 4020 triggers were collected at $B=0 \, $T, 6111 at $0.5 \, $T and 5166 at $B=1\,$T. 
The temperature was 28 to 32 $^\circ$C during the data taking. 
The gas pressure in the vessel followed the atmospheric pressure
which was stable during the runs used in this paper.

In the following analyses, the first two and last two padrows, 
as well as the four leftmost and two rightmost pads of each row, 
are not used in the measurements, to avoid edge effects. 
However, when a hit is found close to the edges of this fiducial region, 
the neighbouring pad(s) in the region are recovered and used in the track reconstruction.

\begin{figure}                                                    
\center                                                          
\epsfig{figure=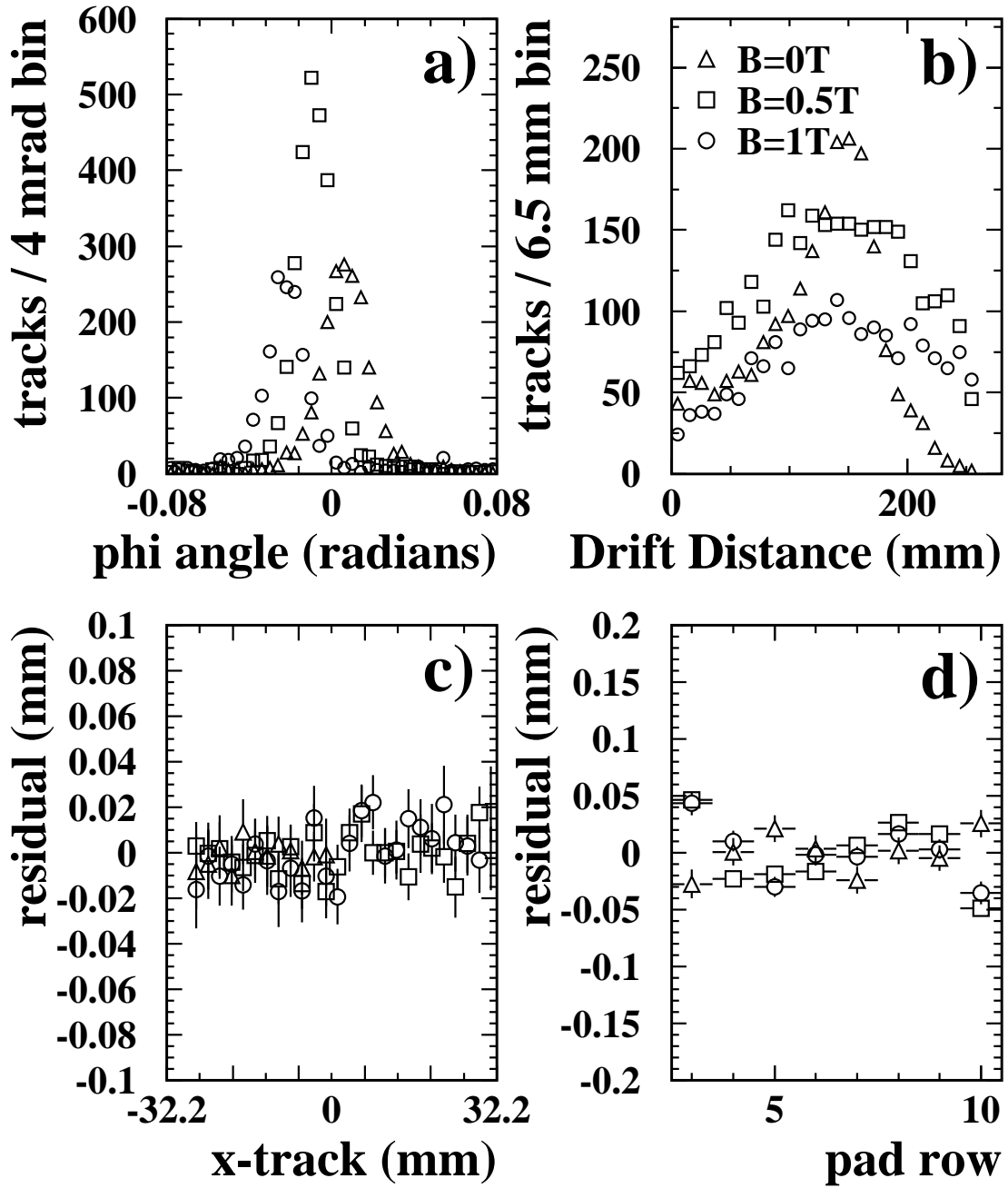,height=20cm} 
\caption{Distribution of the azimuthal angle of the tracks (a) 
and $z$ position at the middle of the tracks (b)
and average residual vs $x$ of the middle of tracks (c) 
and vs padrow number (d) for $B= 0$ (crosses), 0.5 (stars), 
and 1T (circles)
}   
\label{fig:dataqual}
\end{figure}

The azimuthal angle distribution is shown in Fig.~\ref{fig:dataqual}a. 
As stated earlier a  wide beam was obtained 
by adding a $5 \,$cm-thick lead brick in the beam line. 
The resulting $z$ distribution of the tracks is shown in Fig.~\ref{fig:dataqual}b. 
The uniformity of the detector can be assessed from the average residual vs track position 
in $x$ (Fig.~\ref{fig:dataqual}c) and vs padrow number Fig.~\ref{fig:dataqual}d 
for the 8 fiducial padrows. 
Distortions up to 50 microns are observed at non-zero magnetic field 
(likely to be due to $E \times B$ effects), 
but they do not affect the resolution measurements 
where only the r.m.s. of the residuals is considered. 


\subsection{Data Analyses}

The exactly same data sample has been analysed with two independent analysis
methods and compared in the following.
They differ primarily by the way the track parameters are fitted to the pad data in the 
$x$-$y$ plane.
The geometrical track parameters are the intercept of the track with the $y=0$ axis ($x_0$), 
the azimuth at this point ($\phi_0$), and the inverse radius of curvature $(1/R)$.

In the first method, a global maximum-likelihood fit of the geometrical track parameters, 
together with the track width ($\sigma_{\rm track}$) to the charge of the pads is carried out\cite{karlen}. 
In other words,
a track is represented by an arc of a circle with a gaussian charge spread 
of width $\sigma_{\rm track}$ along the $x$ direction, and $\sigma_{\rm track}$ is an additional fit parameter.
The pad charge and time are determined as follows: the ADC counts of the 3 preceding 
and 3 following the time bucket where the maximum is reached are 
added together with the bucket of maximum charge.
This sum is required to exceed a minimum of 7 ADC counts for the pad to be included in a hit. 
The time associated to the hit pad is the charge-weighted average of the seven buckets 
used in the charge integration. 

The log-likelihood function which is maximised with respect to the 4 track parameters
reads:
$\Sigma n_i {\rm ln}(f_i)$, where $f_i$ is the fraction of the
charge expected in pad $i$, 
obtained by integration over the pad of a function with a gaussian profile transverse 
to the track.
$n_i$ is the number of electrons before amplification corresponding 
to the charge readout on the pad $i$. 
This method is implemented in the FORTRAN 95 program FTPC\cite{Ref:Kirsten}.

In the second method, a $\chi^2$ fit is performed to hit points on a row-by-row basis.
The method is implemented in C++ and available as the Double-Fit program\cite{Ref:multifit}.
The Double-fit program starts with cluster finding in the plane of pad-number vs time-bucket for
each pad row.
To build a 2-dimensional ($x$ and $z$) cluster, consecutive time buckets are summed, starting from
the first pulse encountered above $9$ ADC counts, up to the first below 6 counts. Such a hit 
is accepted if its peak pulse height exceeds $15$ ADC counts.
The coordinates of the hit is then calculated as its charge barycenter.
Following the cluster finding, the program then performs a $\chi^2$ fit
of either a straight line at $B=0$ or a circle at $B=0.5$ or $1 \,$T.

\section{Gas Properties}
\subsection{Drift Velocity Measurements}
Measuring the drift velocity of gas mixtures is an interesting test of the electron transport 
simulation in gases,
or alternatively can be considered as a check of gas purity and composition. 
The drift field was set to 220 V/cm
and the detector was taken outside the magnet, itself not energized.
For this measurement a novel and very simple technique was used: 
the beam was shot at an angle of 45 degrees to 
the center of the cathode. 
A $1 \times 1$ cm$^2$ scintillator was added in the trigger to select pions crossing the 
cathode plane in the center. 
Electrons from ionisation close to the cathode drift all the way along the axis of the
TPC reaching the central pads of the detector, 
in the region where no electric-field distortions are expected.
The TPD measures the time elapsed between the trigger arrival and the signal arrival.
The end-point of this time distribution, added to the delay between the trigger and the readout, 
is the time taken by the electrons to drift along the 260.8 mm chamber. 
The drift time is found to be 
$5,907 \pm 30 \, \rm{ns}$.
The trigger delay was measured to be $310 \pm 5 \ \rm{ns}$. 
Adding a rough estimate of 
$30 \pm 20 \, \rm{ns}$ for the NIM to ECL converter at the entrance of the TPD,
and subtracting an estimated $75\,$ns for the pad signal delay in the $15\,$m 
readout cables, 
the total drift time is measured to be $6.172 \pm 0.045~\mu \rm{s}$.
Dividing the chamber length by this time yields  
$v_{\rm drift} = 4.226 \pm 0.031 \, \rm{cm}/ \mu \rm{s}$.
An alternative method has also been used, with a wide beam perpendicular 
to the axis of the chamber and also measuring the 
endpoint of the time distribution. 
It leads to the value of $4.157 \pm 0.036 \, \rm{cm}/ \mu \rm{s}$, 
in agreement with the former. These two values are combined assuming fully correlated systematics
to yield the measurement of the drift velocity of electrons in
Ar+5\% isobutane mixture at an electric field of 220 V/cm of
\begin{eqnarray}\nonumber
v_{\rm drift} & = & 4.181 \pm 0.031 \rm{cm}/ \mu \rm{s}.
\end{eqnarray}
This value is used for the determination of the $z$ coordinate
of the hits along the axis of the chamber in the following data analyses.

This measurement is in very good agreement with the Magboltz\cite{Ref:Biagi} prediction 
of $4.173 \pm 0.016~\rm{cm}/ \mu \rm{s}$.
The uncertainty on the prediction is dominated by the error on the gas composition, 
stemming from a 2\% uncertainty on the isobutane gas flow. 
No consideration on the model, the approximations nor on the input data used in the 
simulation enter this estimate of the prediction uncertainty.

\subsection{Transverse Diffusion Constant Measurements}

For large enough drift distances, so that the ionisation charge is spread over several pads, 
the width of the track
$\sigma_{\rm track}$ allows the determination of the diffusion constant $C_D$ using the relation 
$\sigma_{\rm track} = C_D \sqrt{z}$.

\begin{figure}                                                    
\center                                                          
\epsfig{figure=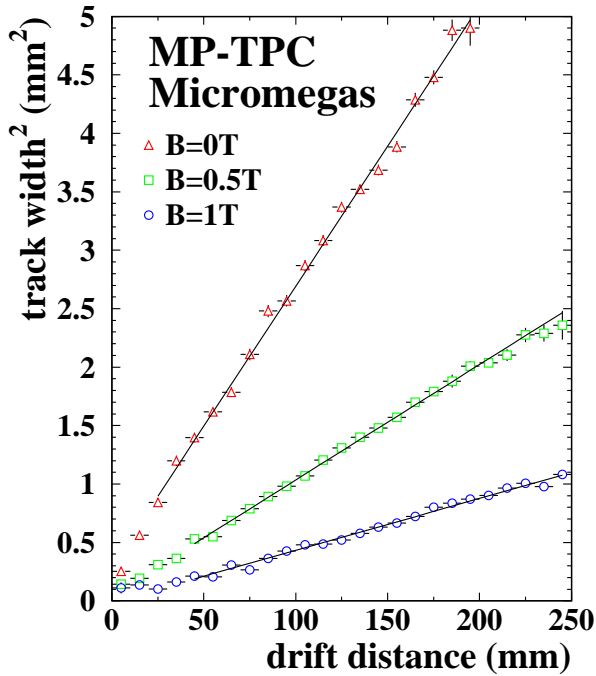,height=10cm} 
\caption{Track width as a function of the drift distance for $B=0$, 0.5, and 1 T}   
\label{fig:diff}
\end{figure}

In the global likelihood method, $\sigma_{\rm track}$ is obtained by maximising the track likelihood simultaneously for
the geometrical track parameters. 
Fig.~\ref{fig:diff} shows for the 3 values of the magnetic field the relation between
the average $\sigma_{\rm track}^2$ and the drift distance $z$. 
The expected linear dependence 
on the drift distance $z$, 
for large $z$, is clearly seen. 
The data at low $z$, where the sensitivity to $\sigma_{\rm track}$ is lost and where the spread in track
widths is comparable to the average track width, are not included into the linear fit. 
The slight offset in the case of zero magnetic field, corresponding to 0.3 mm added 
in quadrature to the track width, can
probably be attributed to delta rays. 
At higher fields the path of these delta-ray electrons 
are expected to be limited by the magnetic field.
The fitted $C_D$ values are given in the first line of Table~\ref{tab:cd}. 
The quoted uncertainties include systematics from the noise level and threshold
used to reconstruct the hits.
  
In the second method, the average fraction of the charge falling on a given pad is plotted 
as a function of the distance
between the track and the pad center. 
It can be shown (see Appendix \ref{prf}) that for large enough drift distances and for high enough
magnetic fields to curl up delta-ray electrons,
this distribution becomes gaussian and its width scales 
as $\sigma_{PR} = \sqrt{ w^2/12 + C_D^2 \, z}$, where $w$ is the pad pitch.
The measured values of $C_D$ by this method are shown in the second line of the table.
These two measurements are in good agreement with each other and 
with the Magboltz prediction, at the level of a few percent. 

\begin{table}[t]
\begin{center}
\caption{Diffusion constants in $\mu \rm{m} / \sqrt{\rm{cm}}$ measured by two methods and Magboltz prediction}

\begin{tabular}{|c|c|c|c|}
\hline \textbf{Magnetic field} & \textbf{0~T} & \textbf{0.5~T} & \textbf{1~T} \\
\hline Global likelihood & $488 \pm 11$ & $314\pm 15$ & $209 \pm 7$ \\
\hline $\chi^2$ method & $475 \pm 5$ & $293 \pm 5$ & $194 \pm 18$ \\
\hline Magboltz & 469.3 & 284.1 & 192.6 \\
\hline
\end{tabular}
\label{tab:cd}
\end{center}
\end{table}


\section{Resolution}


\subsection{Theory and Simulation}

\subsubsection{Effective Number of Ionisation Electrons}
In this section, we present an analytic formula for the spatial resolution
of a TPC equipped with a readout plane consisting of a gaseous
detector that amplifies track electrons and rows of readout pads to measure
their charge centroids.
For simplicity, consider a charged particle passing through the TPC 
at the right angle to a pad row and at a drift distance $z$ from the readout 
plane.
In this zero crossing angle case, as far as the size of the primary ionisation
clusters is negligible\footnote{
This assumption is justifiable at high magnetic field which is expected to
curl up delta rays and hence suppress the size of the primary ionisation 
clusters.
}, 
all of the track electrons created in the ionisation will have 
a $\delta$-function-like distribution peaking at, say, $x=\tilde{x}$,
when projected onto the $x$-axis, namely, 
the axis in the pad row direction.

These track electrons drift toward the pad row.
While drifting,  each of these track electrons experiences 
transverse diffusion and will have an $x$-coordinate deviated from
$x=\tilde{x}$ by $\Delta x_i$, {\it i.e.} $x_i = \tilde{x} + \Delta x_i$,
according to the probability distribution:
\begin{equation}
P_D(\Delta x_i; \sigma_d) 
	= \frac{1}{\sqrt{2\pi}\sigma_d} \, \exp\left(- \frac{\Delta x_i^2}{2\sigma_d^2}\right),
\end{equation}
where the subscript, $i$, means $i$-th electron and $\sigma_d = C_D \sqrt{z}$
with $C_D$ being the transverse diffusion constant.

The number of track electrons that will reach the gas amplification region
in front of the pad row fluctuates statistically. 
Let us denote the probability of getting $N$ such ionisation electrons eventually
contributing to the signal induced on the pad row by $P_I(N; \bar{N})$ with
$\bar{N}$ being the average: $\bar{N} = \left< N \right>$.

Each of these $N$ ionisation electrons will be gas-amplified at the readout
plane by a factor, $G$, which is assumed here to fluctuate according to a Polya distribution:
\begin{eqnarray}
\label{Eq:polya}
P_G(G/\bar{G}; \theta) 
& = & \frac{(\theta + 1)^{\theta+1}}{\Gamma(\theta+1)}
     \left( \frac{G}{\bar{G}} \right)^\theta \exp \left(-(\theta+1)\left(\frac{G}{\bar{G}}\right) \right)
\end{eqnarray}
with $\bar{G}$ being the average gas gain: $\bar{G} = \left< G \right>$.
Notice that the Polya distribution becomes exponential in the $\theta \to 0$ limit,
while it coincides a $\delta$-function in the $\theta \to \infty$ limit. 
Notice also that the Polya distribution has a variance:
$\sigma_{G/\bar{G}}^2 = 1 / (1 + \theta)$, which goes to zero as $\theta \to \infty$
as expected.

For illustration purpose, let us assume, for a while, an idealistic readout plane
which measures the $x$-locations of individual electrons with infinite
accuracy but with relative weights of gas gain values.
Then, the center of gravity of these $N$ electrons at the readout plane will
be given by
\begin{equation}
\label{Eq:cog}
\bar{x} =  \frac{\sum_{i=1}^N G_i \, x_i}{ \sum_{i=1}^N G_i}  
	     =  \tilde{x} + \frac{\sum_{i=1}^N G_i \, \Delta x_i}{ \sum_{i=1}^N G_i} .
\end{equation}

The charge centroid, $\bar{x}$, will then be distributed according to 
\begin{eqnarray}
\label{Eq:pxbarp}
P(\bar{x}; \tilde{x}) & = & \sum_{N=1}^{\infty} P_I(N; \bar{N}) 
				\prod_{i=1}^{N}\left( \int d\Delta x_i P_D(\Delta x_i; \sigma_d) 
				                                       \int d(G_i/\bar{G}) \, P_G(G_i/\bar{G}; \theta) \right) \cr
& ~ & ~~~~~~~~~~~~~~~~~~~~~~~~~~~~~~~~~~~~~~~~~~~~~~~~ \times
\delta\left(\bar{x} - \tilde{x} -  \frac{\sum_{i=1}^N G_i \, \Delta x_i}{ \sum_{i=1}^N G_i} \right) .
\end{eqnarray}

Under the assumption that $N$ is large enough and hence
\begin{equation}
\bar{G} \simeq \frac{1}{N} \sum_{i=1}^N G_i ,
\end{equation}
we obtain the variance of the center of gravity, $\sigma_{\bar{x}}$, 
by inserting Eqs.(\ref{Eq:cog}) and (\ref{Eq:pxbarp}) into its definition and carrying out the integral:
\begin{eqnarray}
\label{Eq:sigxbarp}
\sigma^2_{\bar{x}} & \equiv & \int d\bar{x} \, P(\bar{x}; \tilde{x}) \,  (\bar{x} - \tilde{x})^2 \cr
& \simeq & \sigma_d^2 \left< \frac{1}{N} \right> \left< \left( \frac{G}{\bar{G}}\right)^2  \right>  
\equiv   \sigma_d^2 \frac{1}{N_{\rm eff}},
\end{eqnarray}
where use has been made of
\begin{equation}
\left< \frac{1}{N} \right> = \sum_{N=1}^{\infty} P_I(N; \bar{N}) \frac{1}{N}
\end{equation}
and
\begin{equation}
\left< \left( \frac{G}{\bar{G}}\right)^2  \right> 
= \int \, d(G/\bar{G}) P_G(G/\bar{G}; \theta) \left( \frac{G}{\bar{G}} \right)^2 .
\end{equation}

The $N_{\rm eff}$ is hence given by
\begin{equation}
\label{Eq:Neffp}
N_{\rm eff} =  \frac{1}{  \left< \frac{1}{N} \right> \left< \left( \frac{G}{\bar{G}}\right)^2  \right> }
= \frac{1}{\left< \frac{1}{N} \right>} \left( \frac{1+\theta}{2 + \theta}\right) .
\end{equation}

Notice that $N_{\rm eff}$ is in general significantly smaller than $\bar{N} = \left< N \right>$ due
to ionisation statistics.
The gain fluctuation further reduces $N_{\rm eff}$\cite{kobayashi}
by as much as a factor of two for the exponential gain fluctuation: $\theta = 0$.
\\

In order to find a lower limit on $\theta$, let us consider the total charge 
distribution. 
The total charge after gas amplification is given by
\begin{equation}
Q =  \sum_{i=1}^N G_i 
\end{equation}
and then its probability distribution function by
\begin{equation}
\label{Eq:pQ}
P_Q(Q) = \sum_{N=1}^{\infty} P_I(N; \bar{N}) 
				\prod_{i=1}^{N} \left( \int d(G_i/\bar{G}) \, P_G(G_i/\bar{G}; \theta) \right)
				\delta\left(Q -  \sum_{i=1}^N G_i \right).
\end{equation}
From this we can readily obtain the variation of the total charge as
\begin{eqnarray}
\sigma^2_{Q} & \equiv & \int dQ\, P_Q(Q) \,  Q^2 - \left( \int dQ\, P_Q(Q) \,  Q \right)^2 \cr
& = & \bar{N} \, \bar{G}^2 \, \left( \frac{\sigma_G^2}{\bar{G}^2}+ \frac{\sigma_N^2}{\bar{N}} \right) .
\end{eqnarray}
This implies that
in the case of a Landau-like $P_I(N; \bar{N})$, for which $\bar{N} \ll \sigma_N^2$,
the gas gain fluctuation will not affect the total charge distribution very much.

For X-rays from  $^{55}\rm{Fe}$, however, we expect
\begin{eqnarray}
\left( \sigma_{Q} / \bar{Q} \right)^2  
& = & \frac{1}{\bar{N}} \left(\frac{\sigma_G}{\bar{G}} \right)^2 
          + \left( \frac{\sigma_N}{\bar{N}} \right)^2 \cr \rule{0in}{5.ex}
& = & \frac{1}{\bar{N}} \, \left( \frac{1}{1+\theta}  + F \right),
\end{eqnarray}
where the Fano factor  $F$ is about 0.2 and $\bar{N}$ is about 220 for argon.
The best resolution so far attained with a Micromegas detector with exactly the same type of mesh
is about 6.8\% in r.m.s.\cite{mmFe55},
which implies $\theta \gsim 0.22$. This is a lower limit in the sense that it neglects experimental
effects increasing the width, other than gas gain fluctuations (electronic noise for instance).
However, $\theta$ can be smaller if the Fano factor is less than 0.2. Taking 
a value $F = 0.16$, still consistent with measurements and theoretical calculations~\cite{sbjamboree}, would lead to a
minimum value of 0.166 for $\theta$.

\subsubsection{Effects of Finite Size Pads}

In the last section, we considered gain fluctuation in the gas amplification
process at the end plane detector but assumed that we could measure the
location of a single electron with infinite accuracy.
We now introduce a pad row of pitch $w$ to measure the charge centroid:
\begin{eqnarray}
\bar{x} & = & \sum_{j} Q_j \, (w j) / \sum_{j} Q_{j},
\end{eqnarray}
where $Q_j$ is the charge on pad $j$ and is given as the sum of
contributions from $N$ seed electrons:
\begin{eqnarray}
\label{Eq:Qj}
Q_j = \sum_{i=1}^{N} G_i \cdot f_{j} (\tilde{x} + \Delta x_i) + \Delta Q_j,
\end{eqnarray}
with $f_j$ being the response function of pad $j$ for seed electron $i$ arriving at
the location $\tilde{x} + \Delta x_i$ and $\Delta Q_j$ being the electronic noise
on pad $j$.
Notice that the pad response function is normalized as
\begin{eqnarray}
\sum_{j} f_j  (\tilde{x} + \Delta x_i) & = & 1 .
\end{eqnarray}

The probability distribution for $Q_j$ is then given by
\begin{eqnarray}
\label{Eq:PQj}
P_j(Q_j; \tilde{x}) & = &  \sum_{N=1}^{\infty} P_I(N;\bar{N}) 
			\prod_{i=1}^{N} \left( \int d\Delta x_i P_D(\Delta x_i; \sigma_d) 	 
			                            \int d(G_i/\bar{G}) \, P_G(G_i/\bar{G}; \theta) \right)  \, \cr
&& ~~~~~~~~~~~~
		\times  \,  \int \, d\Delta Q_j \, P_E(\Delta Q_j; \sigma_E)  \,\,
				\delta\left( Q_j - \sum_{i=1}^{N} G_i \cdot f_{j} (\tilde{x} + \Delta x_i) 
		                                  - \Delta Q_j \right),
\end{eqnarray}
where $P_E$ represents a constant electronic noise with 
$\left< \Delta Q_j \right> = 0$ and $\left< \Delta Q_j^2 \right> = \sigma_E^2$.
On the other hand, the probability distribution for the charge centroid is given by
\begin{eqnarray}
\label{Eq:pxbarpp}
P(\bar{x}; \tilde{x}) & = &  \sum_{N=1}^{\infty} P_I(N;\bar{N}) 
			\prod_{i=1}^{N} \left( \int d\Delta x_i P_D(\Delta x_i; \sigma_d) 	 
			                            \int d(G_i/\bar{G}) \, P_G(G_i/\bar{G}; \theta) \right) \cr
&& ~~~~~~~~~~~~~~
		\times \prod_{j}  \left( \int d \Delta Q_j \,\, P_E(\Delta Q_j; \sigma_E)  \,\,
			\int d Q_j \,\,
			\delta\left( Q_j - \sum_{i=1}^{N} G_i \cdot f_{j} (\tilde{x} + \Delta x_i) 
		                                  - \Delta Q_j \right)
					\right) \cr
&& ~~~~~~~~~~~~~~~~~~~~~~~~~~
		\times \delta\left( \bar{x} - \frac{ \sum_{j} Q_j \, (w j)}{ \sum_{j} Q_j} \right)
\end{eqnarray}
which replaces Eq.(\ref{Eq:pxbarp}).

Since the probability distribution $P(\bar{x}; \tilde{x})$ depends on the true location
of the seed cluster $\tilde{x}$, we average over $\tilde{x}$ to define $\sigma_{\bar{x}}$: 
\begin{eqnarray}
\sigma^2_{\bar{x}} 
& \equiv & \int_{-1/2}^{+1/2} d\left(\frac{\tilde{x}}{w}\right)
		\int d\bar{x} \, P(\bar{x}; \tilde{x}) \,  (\bar{x} - \tilde{x})^2 .
\end{eqnarray}
Substituting Eq.(\ref{Eq:pxbarpp}) in this,
and performing integration over
$\Delta Q_j$, $\Delta x_i$, and $\Delta G_i/\bar{G}$ in this order, and then
averaging over $N$, 
we obtain
\begin{eqnarray}
\label{Eq:sigxbarpp}
\sigma_{\bar{x}}^2
& = & \int_{-1/2}^{+1/2} d\left(\frac{\tilde{x}}{w}\right)
		\times \left[ 
				\left( \sum_{j} (j w) \, \left< f_{j} (\tilde{x} + \Delta x) \right> - \tilde{x} \right)^2
				\right. \cr
&& ~~~~~~~~ \left.
				+ \left< \frac{1}{N} \right> \, \left< \left( \frac{G}{\bar{G}} \right)^2 \right> \,
				   \left(
					\sum_{j, k} j k w^2 \, 
						\left< f_{j} (\tilde{x} + \Delta x) f_{k} (\tilde{x} + \Delta x) \right>
					- \left( \sum_{j} j w \left< f_{j} (\tilde{x} + \Delta x) \right> \right)^2
				    \right)
				\right] 
				\cr
&& ~~~~~~~~~~~~~~~~~~~~~~~~~~~~~~~~~~~~~~~~~
				+ \left( \frac{w \sigma_E}{\bar{G}} \right)^2 
						\left< \frac{1}{N^2} \right>  \sum_j j^2 ,
\end{eqnarray}
where we have ignored the electronic noise
as compared to the total charge: 
\begin{eqnarray}
\nonumber
\sum_{j} Q_j
& = & \sum_{i=1}^{N} G_i \, \sum_{j} f_{j} (\tilde{x} + \Delta x_i)
		 + \sum_j \Delta Q_j  
~ \simeq ~ \sum_{i=1}^{N} G_i ,
\end{eqnarray}
and 
\begin{eqnarray}
\nonumber
\sum_{i=1}^{N} G_i & \simeq & N \, \bar{G} 
\end{eqnarray}
as usual.

Notice that the pad response function only appears in the following two forms:
\begin{eqnarray}
\left<  f_{j} ( \tilde{x} + \Delta x) \right>
& \equiv & \int d\Delta x P_D (\Delta x; \sigma_d) \, f_{j} (\tilde{x} + \Delta x)
\end{eqnarray}
and
\begin{eqnarray}
\left<  f_{j} ( \tilde{x} + \Delta x) f_{k} ( \tilde{x} + \Delta x) \right>
& \equiv & \int d\Delta x P_D (\Delta x; \sigma_d) \, f_{j} (\tilde{x} + \Delta x)
\,  f_{k} (\tilde{x} + \Delta x) ,
\end{eqnarray}
and can be numerically evaluated, once the functional form of $f_{j}$ is given.

The formula, Eq.(\ref{Eq:sigxbarpp}), can be qualitatively interpreted as follows.
The first term is the mean square of the difference between
the charge centroid and the true location of the seed cluster and is independent of
the number of primary electrons\footnote{
It is well known that there is an $S$-shape systematics in the difference
between the simple charge centroid and the true cluster location.
This term can hence be eliminated by correcting the charge centroid
for the $S$-shape systematics. 
}.
This term vanishes in the narrow pad limit, $w \to 0$, while it approaches
the famous $(w/\sqrt{12})^2$ in the wide pad limit, $w \gg \sigma_d$.
The second can be interpreted as
the combined effect of the diffusion and
the gas gain fluctuation.
As we have seen in the previous sections, their contributions scale as
$\left< 1/ N \right> \left< (G/\bar{G})^2 \right> \equiv 1/N_{\rm eff}$.
The last term represents the contribution from the electronic noise and is
independent of the shape of the pad response function or the diffusion.
It scales as $ (w \, \sigma_E / \bar{G})^2 \left< 1 / N^2 \right>$.

\subsubsection{Application to a Micromegas-like Readout Plane}

The qualitative observations we made in the last section agree with naive expectations.
For quantitative comparison with data, however, we need a concrete form of
the pad response function.
For simplicity, let us assume that the spatial size of the avalanche
caused by a single seed electron is negligible compared to the pad width,
as expected for a Micromegas-like readout plane.
In this limit, the pad response function becomes hodoscope-like:
\begin{eqnarray}
\label{Eq:prfhodo}
f_{j} (\tilde{x} + \Delta x) 
& \equiv & \Theta\left( (\tilde{x} + \Delta x)/w - j + 1/2 \right) \, 
		\Theta\left( j + 1/2 - (\tilde{x} + \Delta x)/w \right) ,
\end{eqnarray}
for which we have
\begin{eqnarray}
\left< f_{j} ( \tilde{x} + \Delta x) \, f_{k} (\tilde{x} + \Delta x) \right>
& = & \left<  f_{j} ( \tilde{x} + \Delta x)^2 \right> \, \delta_{j k} 
~ = ~ \left<  f_{j} ( \tilde{x} + \Delta x) \right> \, \delta_{j k} .
\end{eqnarray}

Substituting these in Eq.(\ref{Eq:sigxbarpp}), we obtain
\begin{eqnarray}
\label{Eq:sigxbarpph}
\sigma_{\bar{x}}^2 
& \simeq & \int_{-1/2}^{+1/2} d\left(\frac{\tilde{x}}{w}\right)
		\times \left[ 
				\left( \sum_{j} (j w) \, \left< f_{j} (\tilde{x} + \Delta x) \right> - \tilde{x} \right)^2
				\right. \cr
&& ~~~~~~~~ \left.
				+ \left< \frac{1}{N} \right> \, 
					\left< \left( \frac{G}{\bar{G}} \right)^2 \right> \,
				\left( 
					\sum_{j} (j w)^2 \, \left< f_{j} (\tilde{x} + \Delta x) \right> 
				 	- \left( \sum_{j} (j w) \left< f_{j} (\tilde{x} + \Delta x) \right> \right)^2
				\right)
				\right]
				\cr
&& ~~~~~~~~ 
				+ \left( \frac{w \sigma_E}{\bar{G}} \right)^2 
						\left< \frac{1}{N^2} \right>  \sum_j j^2 .
\end{eqnarray}
As long as $\sigma_d \gg w$ as at long distance,
in the integrand of the above equation the first term can be ignored
and the second term can be approximated (see Appendix \ref{scaling}) by
\begin{eqnarray}
\left< \frac{1}{N} \right> \, \left< \left( \frac{G}{\bar{G}} \right)^2 \right> \,
	\left( 
		\sum_{j} (j w)^2 \, \left< f_{j} (\tilde{x} + \Delta x) \right> 
				- \tilde{x}^2
	\right)
& \simeq &  \left< \frac{1}{N} \right> \, 
	\left< \left( \frac{G}{\bar{G}} \right)^2 \right> \,
		\left( \sigma_d^2 + \frac{w^2}{12} \right) .
\end{eqnarray}
In this long drift distance limit, we have an asymptotic formula:
\begin{eqnarray}
\label{Eq:sigxbarpphlongdist}
\sigma_{\bar{x}}^2 
& \simeq & 
		   \left< \left( \frac{G}{\bar{G}} \right)^2 \right> 
			 \,  
			\left< \frac{1}{N} \right> \, \left( \sigma_d^2 + \frac{w^2}{12} \right)
		+ \left( \frac{w \sigma_E}{\bar{G}} \right)^2 
						\left< \frac{1}{N^2} \right>  \sum_j j^2 
\end{eqnarray}
which implies
\begin{eqnarray}
\label{Eq:sigma0}
\sigma_0 
& = & \sqrt{ 
		\left< \frac{1}{N} \right> 
		\left< \left( \frac{G}{\bar{G}} \right)^2 \right> }
		\, 
		\left( \frac{w}{\sqrt{12}} \right)
~ = ~ \left( \frac{w}{\sqrt{12}} \right) \, \frac{1}{\sqrt{N_{\rm eff}}} ,
\end{eqnarray}
if the electronic noise is negligible.

\begin{figure}[h]
\centerline{
\epsfxsize=11cm
\epsfbox{\figdir/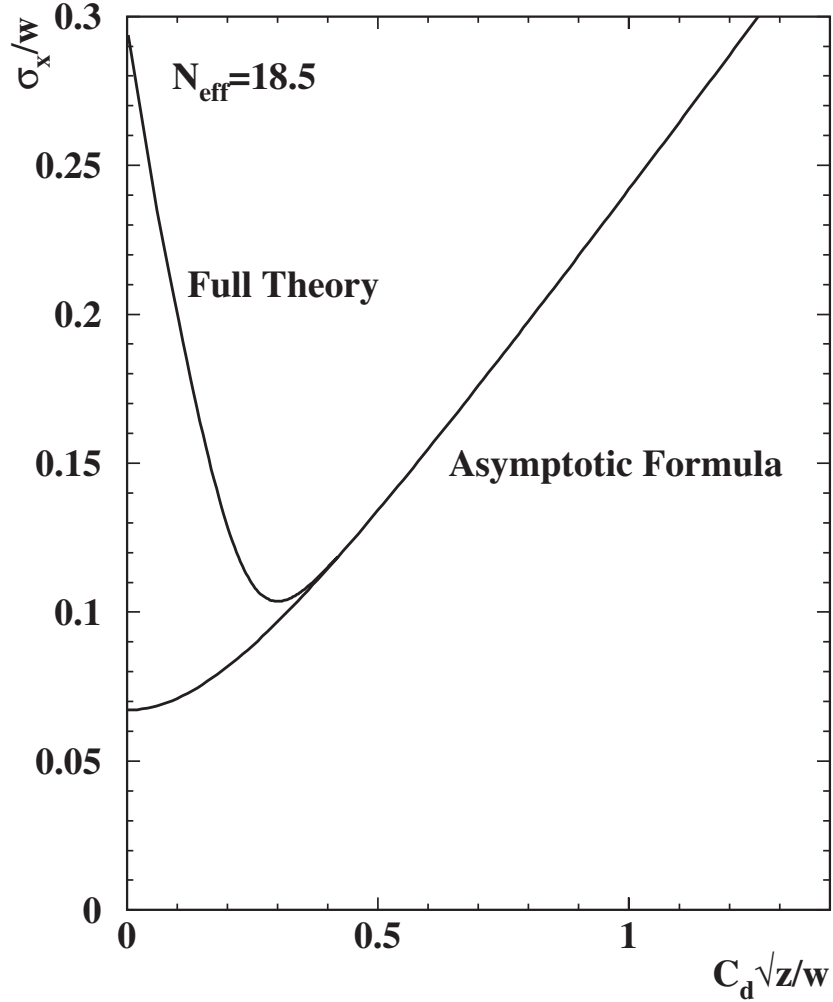}
}
\caption[sgxall]{\label{Fig:sgxscaling}
Expected spatial resolution normalized by the pad pitch for
$N_{\rm eff} = 18.5$ as a function of the scaling variable $\sigma_d / w = C_D \sqrt{z} /  w$.
}
\end{figure}

The integration in Eq.(\ref{Eq:sigxbarpph}) can be carried out numerically at any
drift distance, using
\begin{eqnarray}
\left<  f_{j} ( \tilde{x} + \Delta x) \right>
& \equiv & \int d\Delta x P_D (\Delta x; \sigma_d) f_{j} (\tilde{x} + \Delta x) \cr
& = & \int_{wj - \tilde{x} - w/2}^{wj - \tilde{x} + w/2}  d\Delta x
			\frac{1}{\sqrt{2\pi} \sigma_d} 
				\exp\left(- \frac{1}{2} \left(\frac{\Delta x}{\sigma_d}\right)^2 \right) 
	\cr \rule{0in}{5ex}
& = & {\rm erf} \left(  \frac{(j+1/2) w - \tilde{x}}{\sqrt{2} \sigma_d}  \right) 
	- {\rm erf} \left(  \frac{(j-1/2) w - \tilde{x}}{\sqrt{2} \sigma_d}  \right) .
\end{eqnarray}
This and Eq.(\ref{Eq:sigxbarpph}) imply that $\sigma_{\bar{x}}/w$ is
a function only of $\sigma_{d}/w$ and $N_{\rm eff}$.
Fig.~\ref{Fig:sgxscaling} plots $\sigma_{\bar{x}} / w$ as a function of this
scaling variable, $\sigma_{d} / w$.
Notice that the full theory curve merges into the asymptotic formula at
around $\sigma_d / w \simeq 0.4$, which means that the effect of finite pad pitch
becomes negligible for $\sigma_d / w \gsim 0.4$.
The full theory has a fixed point, $\sigma_{\bar{x}}/w = 1/\sqrt{12}$, at $\sigma_d /w = 0$,
while the asymptotic formula scales as $1/\sqrt{N_{\rm eff}}$.
The full theory curve 
 attains its minimum of about $\sigma_{\bar{x}} / w \simeq 0.1$ 
at around $\sigma_d / w \simeq 0.3$.

\begin{figure}[h]
\centerline{
\epsfxsize=11cm
\epsfbox{\figdir/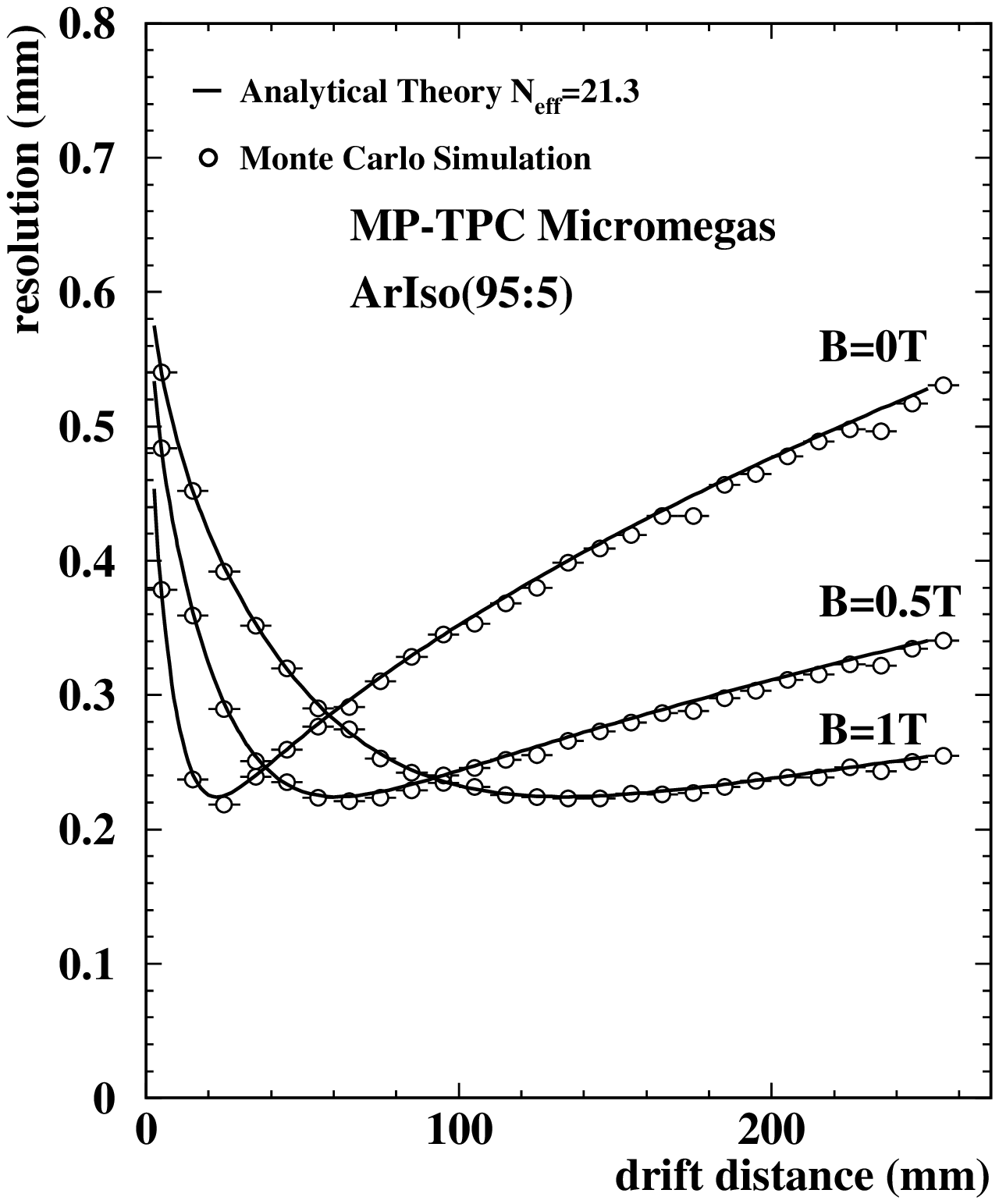}
}
\caption[sgxall]{\label{fig:mccomp}
Expected spatial resolution with readout pads at a pitch of $w = 2.3\,$mm for
$\left< 1/N \right> = 1/38.7$ and $\theta = 0.22$, assuming 
Magboltz results $C_D = 0.469, 0.285$, and $0.193 \,$mm/$\sqrt{\rm cm}$
for $B = 0, 0.5$, and $1.0 \,$T,
respectively.
}
\end{figure}

\subsubsection{Comparison with Monte Carlo Simulation}

A Monte-Carlo simulation has been carried out to check the analytical theory and to estimate 
the expected effective number of electrons for the gas mixture used in the tests. 

In a first step track segments are generated with a uniform distribution of $x$ across a 2.3~mm pad 
and for a set of $z$ values along the drift axis. 
The number of ionisation clusters 
is generated along a 6.3 mm pad according to a Poisson distribution with mean value 32.0. 
This average number of clusters is estimated as follows: a concentration-weighted average of 
the number of clusters from minimum ionizing particles (m.i.p.)
in Ar (23 e$^-$/cm) and in isobutane (84 e$^-$/cm), taken from ref.~\cite{schmidt}, 
multiplied by the calculated 
ratio of ionisation $dN/dx$ for 4 GeV pions with respect to m.i.p., taken to be 1.23 from ref.~\cite{santovetti}.
In each cluster a number of electrons is generated according to 
the argon cluster size distribution given in ref.~\cite{fischle}. 
The average $1/N$ is predicted to be 1/38.7. 

Then each individual electron is transported over the distance $z$. 
The transverse diffusion in the $x$
direction is simulated by varying $x$ by a random amount, following a gaussian law, 
the width of which is given by $C_D \sqrt{z}$, 
where $C_D$ is the Magboltz prediction for the diffusion constant : 
$C_D = 0.469, 0.285$, and $0.193 \, $mm/$\sqrt{\rm cm}$ for $B = 0, 0.5$, and $1.0 \,$T,
respectively.

The last step is to simulate the gas amplification gain. For this a Polya distribution with the 
$\theta$ parameter equal to 0.22 is used. For every track, the charge sum on each pad is calculated.
Then hits are reconstructed as barycentres of the pads hit.
The $x$ resolution for each $z$ is plotted in Fig.~\ref{fig:mccomp}. 
The analytical formula, with input $N_{\rm eff} = 21.3$, as given by Eq.~\ref{Eq:Neffp}, 
reproduces the Monte Carlo data very well.


\subsection{Comparison with Measurements}
  
\begin{figure}                                                    
\center
\begin{minipage}[htb]{17cm}
\epsfig{figure=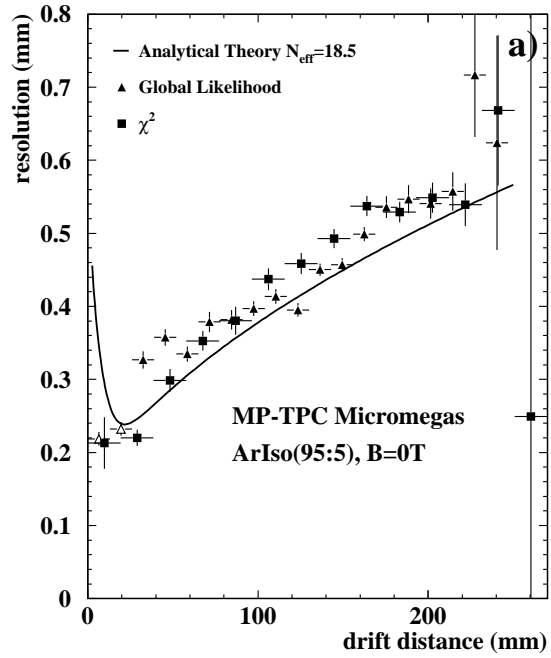,height=10cm}
\epsfig{figure=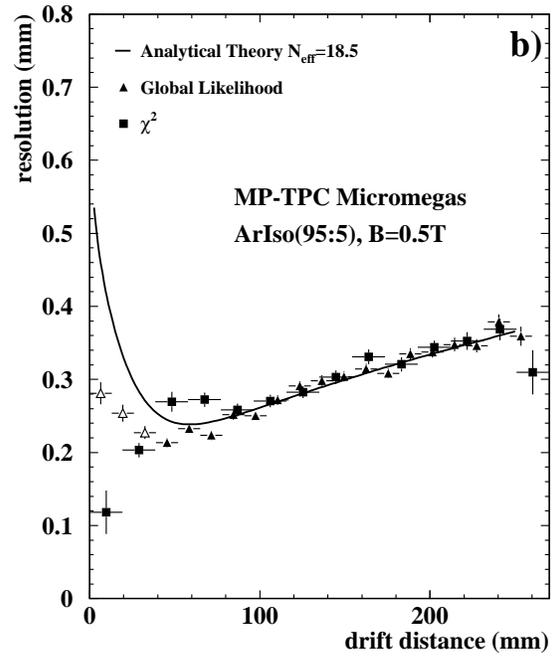,height=10cm} 
\epsfig{figure=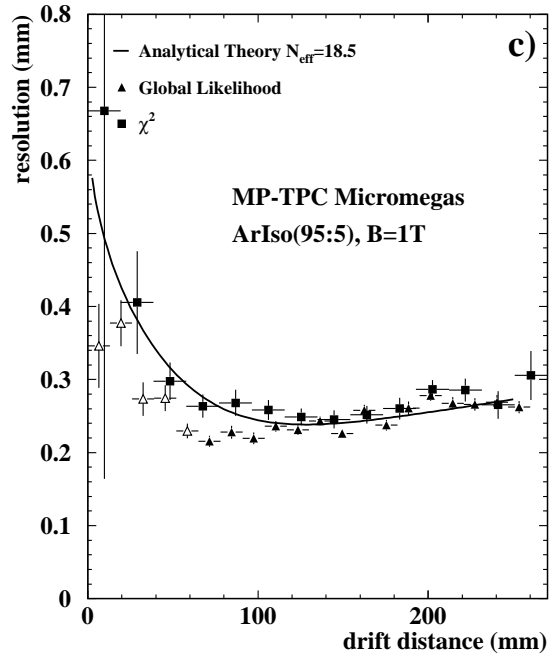,height=10cm} 
\end{minipage}
\caption{
Resolutions at (a) $B=0$, (b) $B=0.5$, and (c) $B=1$~T, as a function
of the drift length
}
\label{fig:resol}
\end{figure}

In the global likelihood analysis, the hit position on each row is estimated by a fit to the charge depositions in the row, with the track parameters fixed to their value 
using the 8 fiducial rows.
The resolution is then calculated as the geometric mean of the r.m.s. values of 
the distributions of the residuals with and without the pad row in question in the fit (see Appendix \ref{geomean})
so as to eliminate contributions from the tracking errors.
To avoid being sensitive to 
outliers (produced by noise for instance) the hits situated 
at a distance of more than 4 standard deviations from the 
track are ignored. 
Alternatively, a gaussian function was fitted to the whole residual distribution 
to estimate its  r.m.s.. 
This second estimate of the resolution is lower than the former by only 7\%. 
The resolutions at $B=0, 0.5,$ and $1 \,$T are shown as a function of the drift distance
in Figs.~\ref{fig:resol}a), b), and c), respectively. 
The results of the global
likelihood method are the triangles. 
The $\chi^2$ method results are also shown (square data points).
The two methods agree each other very well at long drift distances. 
The discrepancies at short distances will be discussed later.
The likelihood method makes a better use of the pad signal information. 

These measured resolutions are compared with the theoretical predictions 
explained in the previous section, with $N_{\rm eff}=18.5$. 
The $\chi^2$ fit using the barycentre of each row leads to values 
of $N_{\rm eff}$ consistent with this (see Table~\ref{tab:neff}).
The theory reproduces the data very well when $C_D \sqrt{z} / w \gsim 0.4$
as expected for $B=0.5$ and $1 \,$T.
The theory seems to underestimate the resolution even at longer distances
at $B=0$ and the discrepancy seems independent of the drift distance.
This discrepancy  could be
attributed to the finite geometrical size of primary ionisation clusters
that would be significant only when
the curling up effect of delta-ray electrons is absent.

\begin{figure}                                                    
\center                                                          
\epsfig{figure=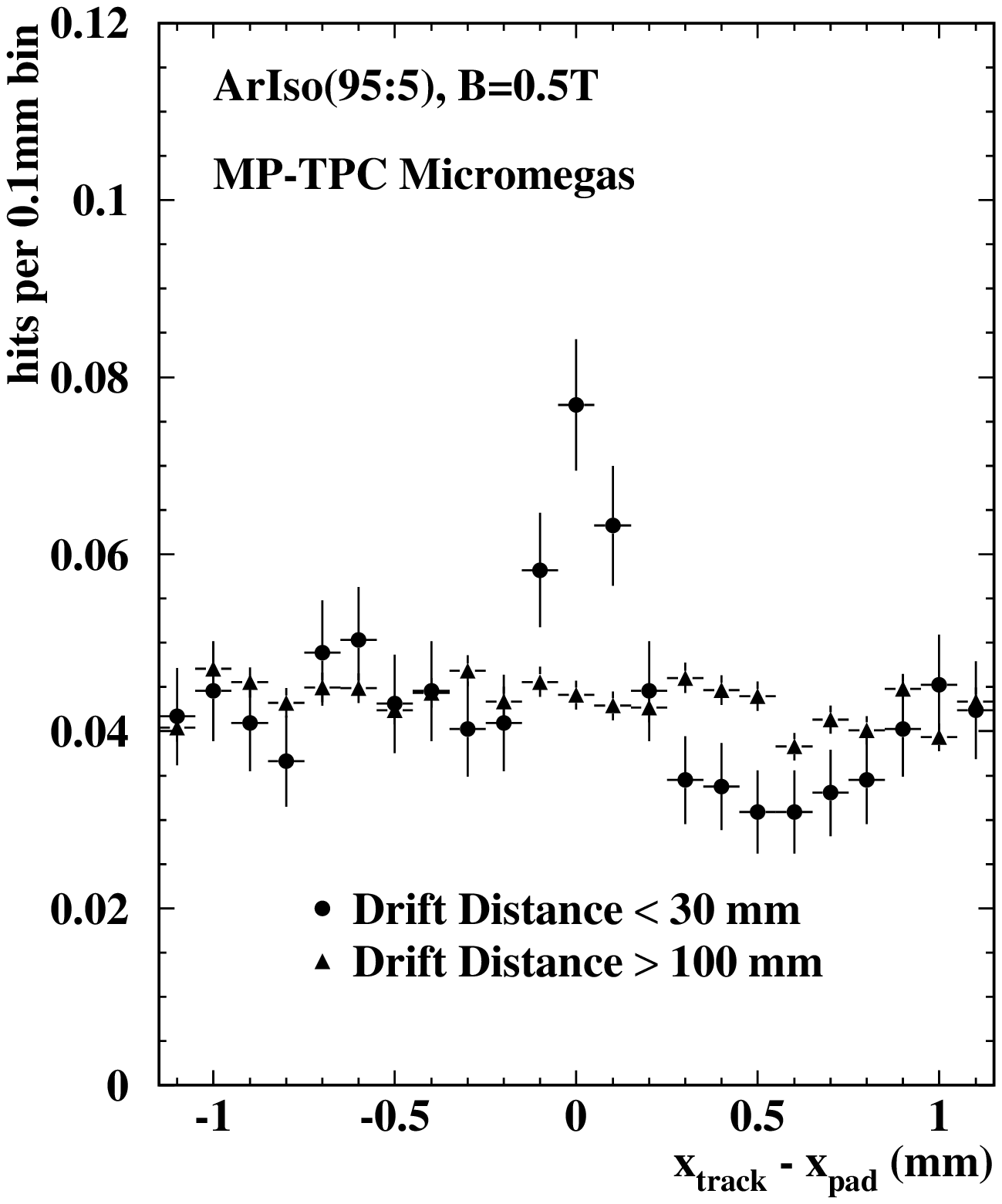,height=12cm}
\caption{Distribution of the distance between the track and the center of a pad for a small 
drift distance sample (dots) and a large drift distance sample (triangles) from 
the global likelihood analysis.}   
\label{fig:bias}
\end{figure}

The hollow data points at small drift distance are not used in the fit. 
The reason to discard them is that they are biased toward low values, 
as the hits are reconstructed towards the middle of the pads. 
An example of this is given
in Fig.~\ref{fig:bias}, 
where the distribution of the position of hits with respect to the middle of a pad is shown; 
it is clear from this plot that for too small drift distance the hits are preferentially reconstructed 
in the middle of a pad, 
so that the whole track is biased towards the middle of the pad 
and the residuals are underestimated. 
This is the main reason why the rise of the resolution at low $z$  
is underestimated in the data (this effect is less prominent in the case of $1 \,$T where 
the track curvature in the magnetic field plays the role of an effective pad staggering).
This effect is present in both methods and could be avoided only if we had had 
an external measurement of the track position.
Notice, however, that unlike the $\chi^2$ method which uses the charge barycentres,
the likelihood method is free from the $S$-shape systematics
(the first term in Eq.~\ref{Eq:sigxbarpp}) 
as long as each pad row has multiple pads above threshold.
This is why the likelihood method tends to deviate from the theory
and give better resolutions than the $\chi^2$ method in the short drift distance region. 

The fitted values of $N_{\rm eff}$ are given in Table~\ref{tab:neff}. Considering that $N_{\rm eff}$ 
is independent of the magnetic field, 
the 3 data sets can be combined to obtain a measurement of $N_{\rm eff} = 18.5 \pm 1.1$.
This value can be compared with the expectation $N_{\rm eff} = 21.3 \pm 2.7$ obtained in 
the previous section fixing the gain fluctuation parameter $\theta$ to 0.22, where the error is evaluated by
varying all the other input parameters 
within their admissible range and adding in quadrature the variations of $N_{\rm eff}$.
The agreement is good, and favors significant gas gain fluctuations.

As explained in the previous section, a non-zero value of 
$\sigma_0$ is expected from the
fit to the data points of the functional form 
\begin{eqnarray}
\nonumber
  \sigma = \sqrt{\sigma_0 ^2 + C_D^2 z / N_{\rm eff}} . 
\end{eqnarray}
The combined measurement of $\sigma_0$ is $165 \pm 18 \, \mu \rm{m}$, 
in good agreement with the expectation 
$w/ \sqrt{12 N_{\rm eff}} = 154 \, \mu \rm{m}$.

\begin{table}[t]
\begin{center}
\caption{Effective number of electrons $N_{\rm eff}$ measured by the two methods}

\begin{tabular}{|c|c|c|c|}
\hline \textbf{Magnetic field} & \textbf{0~T} & \textbf{0.5~T} & \textbf{1~T} \\
\hline Global likelihood & $16.5 \pm 3.4$ & $18.1 \pm 1.2$ & $22.8 \pm 3.0$ \\
\hline $\chi^2$ fit + barycentre & $15.1 \pm 1.2$ & $18.7 \pm 2.6$ & $15.7 \pm 7.7$ \\
\hline
\end{tabular}
\label{tab:neff}
\end{center}
\end{table}

  \subsection{Extrapolation to ILC-TPC}
\begin{figure}                                                    
\center                                                          
\epsfig{figure=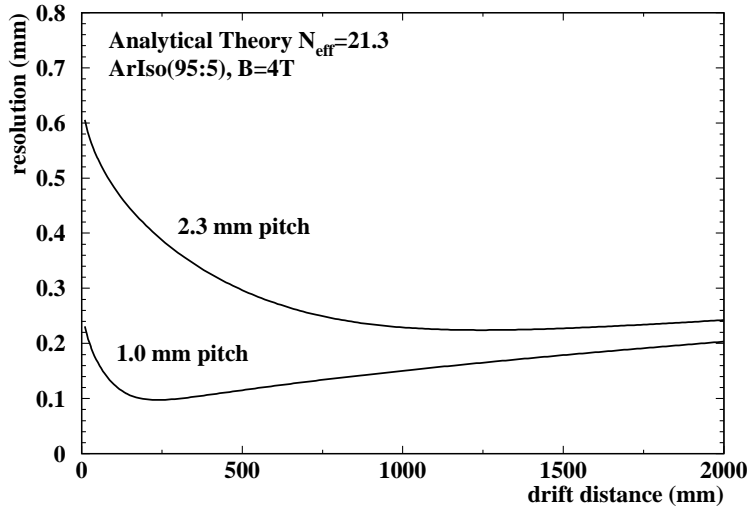,height=8cm} 
\caption{Resolution in the Linear Collider case as a function of the drift length, for two values of the pitch,
for the gas used in this study}   
\label{fig:extrapiso}
\end{figure}

Conforted that the theory of resolution presented here is a pertinent approximation for 
a Micromegas TPC, we can use it to predict what can be expected in the 4 Tesla case, 
as proposed for the Linear Collider TPC. 
Fig.~\ref{fig:extrapiso} represents the expected resolution
as a function of $z$, for two values of the pitch for the gas studied in this paper.
One can see that at such a high magnetic field, the
diffusion, with a constant of 63 $\mu$m/$\sqrt{\rm cm}$, is not sufficient 
to spread the charge enough.
This results in the fact 
that the hodoscope effect is felt up to distances over one meter in the case of a 2.3~mm pitch.
Even a 1~mm pitch, which would require a very dense readout electronics 
and would feature pads with a large 
aspect ratio, would not allow the target average resolution of 100 $\mu$m to be reached.

\begin{figure}                                                    
\center                                                          
\epsfig{figure=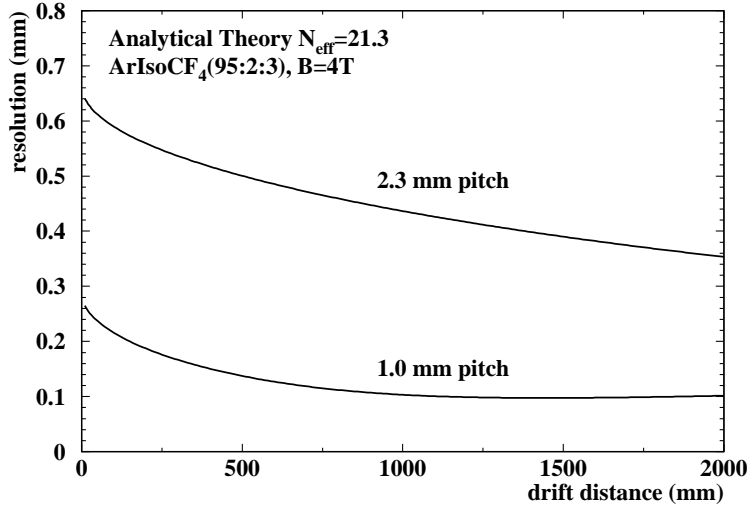,height=8cm} 
\caption{Resolution in the Linear Collider case as a function of the drift length, for two values of the pitch,
for a fast gas mixture with a large collision time.}   
\label{fig:extrapmix}
\end{figure}

In the case of the triple mixture Ar:isobutane:CF$_4$ (95:2:3), with a record diffusion constant 
as low as 26 $\mu$m/$\sqrt{\rm cm}$, 
the situation is even more catastrophic: the hodoscope effect is felt over all distances
even with 1~mm pads (Fig.~\ref{fig:extrapmix}).

This study shows that, to obtain the target resolution required to fulfill the ILC physics program, 
either smaller, digital pads are necessary, 
or a spreading of the charge onto several pads has to be implemented
after amplification. 
The latter can be achieved either in a multi-GEM structure by maximizing the natural
defocusing in the transfer between two GEMs, 
or by the use of a resistive-capacitive continuous layer
(resistive foil) as proposed by some of us~\cite{dixit}.


\section{Conclusions}
    Successful operation of a Micromegas TPC in a test beam enabled us to 
measure its spatial resolution under a magnetic field as well as the properties	
of an Ar isobutane mixture. 
The drift velocity and the transverse diffusion constant for at $B=0, 0.5$ and $1 \,$T 
are found to be in good agreement with the predictions of
Magboltz. The obtained spatial resolution as a function of drift distance is well reproduced by an analytical
calculation developed in this paper. 
The effective number of electrons $(N_{\rm eff})$ is measured to be $18.5 \pm 1.1$, 
which is consistent with an estimate based on the primary ionisation statistics and
the gas gain fluctuations.

The extrapolation to the conditions of the Linear Collider shows that the goal resolution 
of 100 microns cannot 
be reached with $2.3 \,$mm pads with the technique used in this test, 
and that another technique will probably be 
necessary (digital smaller pad readout or resistive anode readout, for instance).

\section*{Acknowledgments}
We would like to thank J. Pouthas
for his kind support. We recognize D. Karlen for his help in using his JTPC display program 
and M.E. Janssen for his Doublefit analysis program.
External institutes wish to thank the
KEK directorate for the hospitality 
extended to them during the beam tests. We  also thank the
Cryomagnetism department for the operation of the magnets.

\appendix

\section{Pad Response Function in the Large $\protect\mathbold{\sigma_{d}}$ limit}
\label{prf}

From Eqs. (\ref{Eq:Qj}) and (\ref{Eq:PQj})
we obtain the average charge on pad $j$:
\begin{eqnarray}
\nonumber
\left< Q_j (\tilde{x}) \right>
& = & \bar{N} \bar{G} \, \left<  f_{j} ( \tilde{x} + \Delta x) \right> ,
\end{eqnarray}
resulting in the average charge fraction on pad $j$:
\begin{eqnarray}
\nonumber
\left< Q_j (\tilde{x}) \right> / (\bar{N} \bar{G})
& = & \left<  f_{j} ( \tilde{x} + \Delta x) \right>
	~ \equiv ~ \int d\Delta x P_D (\Delta x; \sigma_d) f_{j} (\tilde{x} + \Delta x)
	\cr \rule{0in}{5ex}
& = & \int_{wj - \tilde{x} - w/2}^{wj - \tilde{x} + w/2}  d\Delta x
			\frac{1}{\sqrt{2\pi} \sigma_d}
				\exp\left(- \frac{1}{2} \left(\frac{\Delta x}{\sigma_d}\right)^2 \right)
	\cr \rule{0in}{5ex}
& = & \int_{- w/2}^{+ w/2}  d\xi
	\frac{1}{\sqrt{2\pi} \sigma_d}
		\exp\left(- \frac{1}{2} \left(\frac{jw + \xi - \tilde{x}}{\sigma_d}\right)^2 \right)  .
\end{eqnarray}
In our standard pad response function analysis, we plot this as a function of
the pad center as measured from the average charge centroid:
\begin{eqnarray}
\nonumber
\hat{x}(\tilde{x})
& \equiv & jw - \left< \bar{x}(\tilde{x}) \right>
~ \equiv ~ jw - \sum_k \, (kw) \left< f_k (\tilde{x} + \Delta x) \right>
	\cr  \rule{0in}{3ex}
& = & jw - \tilde{x} + O \left( \left(\frac{w}{\sigma_d} \right)^2 \right)
~ \simeq ~  jw - \tilde{x}.
\end{eqnarray}
In the large $\sigma_d$ limit, the charge fraction hence has the following
functional form:
\begin{eqnarray}
Q_{PR} (\hat{x})
& \simeq &  \frac{1}{w} \,
	\int_{- w/2}^{+ w/2}  d\xi
	\frac{1}{\sqrt{2\pi} \sigma_d}
		\exp\left(- \frac{1}{2} \left(\frac{\hat{x} + \xi}{\sigma_d}\right)^2 \right)  .
\end{eqnarray}
Since $Q_{PR}(\hat{x})$ is apparently normalized to unity, we have
\begin{eqnarray}
\nonumber
\sigma_{PR}^2
& = & \int_{-\infty}^{+\infty} d\hat{x} \, Q_{PR} (\hat{x}) \, \hat{x}^2
	\cr
& \simeq & \frac{1}{w} \, \int_{- w/2}^{+ w/2}  d\xi \,
	\int_{-\infty}^{+\infty} d\hat{x} \,
	\frac{1}{\sqrt{2\pi} \sigma_d}
		\exp\left(- \frac{1}{2} \left(\frac{\hat{x} + \xi}{\sigma_d}\right)^2 \right) 	
	\hat{x}^2
	\cr \rule{0in}{5ex}
& = & \frac{1}{w} \,  \int_{- w/2}^{+ w/2}  d\xi \,
	(\sigma_d^2 + \xi^2)
~ = ~ \sigma_d^2 + \frac{w^2}{12} ,
\end{eqnarray}
and, consequently,
\begin{eqnarray}
\sigma_{PR}^2 (0)
& \equiv & \lim_{\sigma_d \to \infty} ( \sigma_{PR}^2 - \sigma_d^2 )
~ = ~ \frac{w^2}{12} .
\end{eqnarray}

\section{Scaling and $\protect\mathbold{\sigma_{d} \to \infty}$ Limit}
\label{scaling}

As long as the avalanche can be regarded as point-like, and the pad response
function is hodoscope-like as given by Eq.(\ref{Eq:prfhodo}),
any observable with the dimension of length should scale as
$\sigma_d$ times a function of  $(w/\sigma_d)$ or
$w$ times a function of $(w/\sigma_d)$.
In the case of the first term of Eq.(\ref{Eq:sigxbarpph}), it should hence scale as
\begin{eqnarray}
\int_{-1/2}^{+1/2} d\left(\frac{\tilde{x}}{w}\right)
	\left( \sum_{j} (j w) \, \left< f_{j} (\tilde{x} + \Delta x) \right> - \tilde{x} \right)^2
& = & \left[ \sigma_d \, F_1(\sigma_d/w) \right]^2,
\end{eqnarray}
where the pad pitch ($w$) can only appear in the function through the ratio: $\sigma_d/w$.
This term represents the well known $S$-shape systematic bias
in the charge centroid for a finite pad pitch and will vanish in the $w \to 0$ limit:
$F_1(\infty) = 0$.
It is, however, non-trivial whether this will vanish in the $\sigma_d \to \infty$ limit
for a fixed pad pitch: $w = \mbox{constant}$.
We can show analytically that this is indeed the case, as follows:
\begin{eqnarray}
\nonumber
\sum_{j} \, (j w) \, \left< f_{j} (\tilde{x} + \Delta x) \right>
& = & \sum_{j} \,  (j w) \,
		\int_{(j-1/2)\,w - \tilde{x}}^{(j+1/2)\,w - \tilde{x}}
			d\Delta x \, P_D (\Delta x; \sigma_d)
	\cr
& = & \sum_{j=1}^{\infty} \, (j w) \,
	\left[ \int_{(j-1/2)w - \tilde{x}}^{(j+1/2)w - \tilde{x}}
		- \int_{(-j-1/2)w - \tilde{x}}^{(-j+1/2)w - \tilde{x}} \right] d\Delta x \,
		P_D(\Delta x; \sigma_d)
	\cr
& = & \sum_{j=1}^{\infty} \, (j w) \,
	\left[ \int_{(j-1/2)w - \tilde{x}}^{(j+1/2)w - \tilde{x}}
		- \int_{(j-1/2)w + \tilde{x}}^{(j+1/2)w + \tilde{x}} \right] d\Delta x \,
		P_D(\Delta x; \sigma_d)
	\cr
& = & \sum_{j=1}^{\infty} \, w \,
	\left[ \int_{(j-1/2)w - \tilde{x}}^{\infty}
		- \int_{(j-1/2)w + \tilde{x}}^{\infty} \right] d\Delta x \,
		P_D(\Delta x; \sigma_d)
	\cr
& = & \sum_{j=1}^{\infty} \, w \,
		\int_{(j-1/2)w - \tilde{x}}^{(j-1/2)w + \tilde{x}}  d\Delta x \,
		P_D(\Delta x; \sigma_d) ,
\end{eqnarray}
where we have used the fact that $P_D(\Delta x; \sigma_d)$ is an even
function of $\Delta x$.

Noting that $-w/2 \le \tilde{x} \le +w/2$
and hence $|\tilde{x}| \ll \sigma_d$ in the large $\sigma_d$ limit,
we can further the calculation by Taylor expansion:
\begin{eqnarray}
\nonumber
\sum_{j} \, (j w) \, \left< f_{j} (\tilde{x} + \Delta x) \right>
& = & \sum_{j=1}^{\infty} \, w \,
		\int_{(j-1/2)w - \tilde{x}}^{(j-1/2)w + \tilde{x}}  d\Delta x \,
		P_D(\Delta x; \sigma_d)
	\cr
& = & \sum_{j=1}^{\infty} \, w \,
		\int_{-\tilde{x}}^{+\tilde{x}}  d\xi \,
		P_D((j-1/2)w + \xi; \sigma_d)
	\cr
& = & \sum_{j=1}^{\infty} \, w \,
		\int_{-\tilde{x}}^{+\tilde{x}}  d\xi \,
		P_D((j-1/2)w; \sigma_d)
		\left(
			1 - \frac{(j-1/2)w}{\sigma_d^2} \, \xi
			+ O\left( \left(\frac{\xi}{\sigma_d}\right)^2\right)
		\right)
	\cr
& \simeq & \sum_{j=1}^{\infty} \, w \,
		P_D((j-1/2)w; \sigma_d) \,
		\int_{-\tilde{x}}^{+\tilde{x}}  d\xi \,
		\left(
			1 - \frac{(j-1/2)w}{\sigma_d^2} \, \xi
		\right)
	\cr
& \simeq & 2 \tilde{x} \, \sum_{j=1}^{\infty} \, w \,
		P_D((j-1/2)w; \sigma_d)
	\cr
& = & 2 \tilde{x} \, \sum_{j=1}^{\infty} \,
		\int_{-w/2}^{+w/2}  d\xi \,
		P_D((j-1/2)w; \sigma_d)
		\left(
			1 - \frac{(j-1/2)w}{\sigma_d^2} \, \xi
			+ O\left( \left(\frac{\xi}{\sigma_d}\right)^2\right)
		\right)
	\cr
& \simeq & 2 \tilde{x} \, \sum_{j=1}^{\infty} \,
		\int_{-w/2}^{+w/2}  d\xi \,
		P_D((j-1/2)w + \xi; \sigma_d)
	\cr
& = & 2 \tilde{x} \int_{0}^{\infty} d\Delta x \, P_D(\Delta x; \sigma_d)
~ = ~ \tilde{x} .
\end{eqnarray}


In order to see the large $\sigma_d$ limit of the second and the third terms
of Eq.(\ref{Eq:sigxbarpph}), we need to evaluate the following:
\begin{eqnarray}
\label{Eq:I}
I & \equiv &
	\sum_{j} \, (j w)^2 \, \left< f_{j} (\tilde{x} + \Delta x) \right>  - \tilde{x}^2 \cr
& = & \sum_{j} \, (j w)^2 \,
		\int_{(j-1/2)\,w - \tilde{x}}^{(j+1/2)\,w - \tilde{x}}
			d\Delta x P_D (\Delta x; \sigma_d)
		- \tilde{x}^2
		\cr
& = & \sum_{j} \,
		\int_{-w/2}^{+w/2} d \xi \,
			\left[ (j w + \xi)^2 - 2 (j w) \xi - \xi^2 \right]
			 P_D (j w  + \xi - \tilde{x}; \sigma_d)
		- \tilde{x}^2
		\cr
& = & \sum_{j} \,
		\int_{-w/2}^{+w/2} d \xi \,
			(jw + \xi)^2
			 P_D (j w  + \xi - \tilde{x}; \sigma_d) - \tilde{x}^2
		\cr
&&  ~~ - \sum_{j} \,
		\int_{-w/2}^{+w/2} d\xi \,
			\left[ 2(jw) \, \xi  + \xi^2 \right]
			 P_D (j w  + \xi- \tilde{x}; \sigma_d)
		\cr
& = & \int  d x \, x^2 \, P_D (x - \tilde{x}; \sigma_d) - \tilde{x}^2
				\cr
&& ~~ -  \sum_{j} \,  P_D (j w - \tilde{x}; \sigma_d)
		\int_{-w/2}^{+w/2} d\xi \,
			\left[ 2(jw) \, \xi  + \xi^2 \right]
			\left(
				1 - \frac{j w - \tilde{x}}{\sigma_d^2} \, \xi
				+ O\left( \left(\frac{\xi}{\sigma_d}\right)^2\right)
			\right) \cr \rule{0in}{3ex}
& \simeq & \sigma_d^2
-  \sum_{j} \,  P_D (j w - \tilde{x}; \sigma_d)
		\left[ 2 \left( (jw) \tilde{x} - (jw)^2 \right) + \sigma_d^2 \right] \,
		\int_{-w/2}^{+w/2} d\xi \,
			\left(\frac{\xi}{\sigma_d}\right)^2  \cr
& = & \sigma_d^2
-  \frac{w^2}{12 \sigma_d^2} \, \sum_{j} \,
		\left[ 2 \left( (jw) \tilde{x} - (jw)^2 \right) + \sigma_d^2 \right] \,
		P_D (j w - \tilde{x}; \sigma_d) \, w \cr
& \simeq & \sigma_d^2
-  \frac{w^2}{12 \sigma_d^2} \, \sum_{j} \,
		\left[ 2 \left( (jw) \tilde{x} - (jw)^2 \right) + \sigma_d^2 \right] \,
		\int_{-w/2}^{+w/2}  d\xi \, P_D (j w + \xi - \tilde{x}; \sigma_d)  \cr
& \simeq & \sigma_d^2
	- \frac{w^2}{12 \sigma_d^2} \, \left[ 2 \tilde{x}^2 - 2 I + \sigma_d^2 \right]
~ \simeq ~  \sigma_d^2
	+ \frac{w^2}{12 \sigma_d^2} \, \left[ 2 I - \sigma_d^2 \right]
\end{eqnarray}
where use has been made of Taylor expansion to the first order of
$\xi$:
\begin{eqnarray}
\nonumber
P_D (j w + \xi - \tilde{x}; \sigma_d)
& \simeq & P_D (j w -  \tilde{x}; \sigma_d)
		- P_D (j w -  \tilde{x}; \sigma_d) \left( \frac{jw - \tilde{x}}{\sigma_d^2}\right) \, \xi
		\cr
& = &  P_D (j w -  \tilde{x}; \sigma_d) \left( 1 - \frac{jw - \tilde{x}}{\sigma_d^2} \, \xi \right)
\end{eqnarray}
and the fact that the odd functions of $\xi$ vanishes upon integration and $\tilde{x}^2 \le w^2/4$.
Substituting the right hand side of Eq.(\ref{Eq:I}) in Eq.(\ref{Eq:I}) iteratively
and ignoring the terms of $O((w^2/\sigma_d^2)^2)$, we finally arrive at
\begin{eqnarray}
I & \simeq &
\sigma_d^2  + \frac{w^2}{12 \sigma_d^2}
\left[ 2 \sigma_d^2 - \sigma_d^2 \right]
~ \simeq ~ \sigma_d^2 + \frac{w^2}{12} .
\end{eqnarray}


Notice  that $\sigma_d$-independent term can be
regarded as proportional to $\left(\sigma_d \times (w/\sigma_d)\right)^2$, and hence
is no exception for the scaling law.

%

\section{Geometric Mean Method}
\label{geomean}

In this appendix we give a simple demonstration of the geometric mean method
applied in the analysis to estimate the spatial resolution.

First, in the case where the hit point in question 
(say, the $i$-th point $x_i$) 
is {\it excluded\/} in the track fitting,
the residual is given by 
\begin{displaymath}
 \Delta x_i = x_i - \hat{x}_i\;,
\end{displaymath}
where $\hat{x}_i$ represents the estimator for the $i$-th point
given by the track fitting using the remaining hit points.
Its variance is
\begin{equation}
\sigma_{\rm excl}^2 \equiv \left<(\Delta x_i)^2\right> =
                       \sigma_{x_i}^2 + \sigma_{\hat{x}_i}^2 \label{a1}\;,  
\end{equation}   
the sum of the {\it true\/} spatial resolution and the tracking error.

Next, in the case where the hit point in question is {\it included\/}
in the track fitting,
the estimator for the $i$-th hit point is given by the weighted mean of
$\hat{x}_i$ and $x_i$:
\begin{displaymath}
 \hat{x}_i^\prime = \frac{w_{\hat{x}_i} \hat{x}_i + w_{x_i} x_i}
                        {w_{\hat{x}_i} +  w_{x_i}} \;,
\end{displaymath}   
with $w_{\hat{x}_i}$ $(w_{x_i})$ being the corresponding weight:
$1/{\sigma_{\hat{x}_i}^2}$ $(1/{\sigma_{x_i}^2})$.
The residual is hence given by 
\begin{displaymath}
 \Delta x_i^\prime \equiv x_i - \hat{x}_i^\prime = 
    \frac{\sigma_{x_i}^2}{\sigma_{x_i}^2 + \sigma_{\hat{x}_i^2}}
     (x_i - \hat{x}_i)
  = \frac{\sigma_{x_i}^2}{\sigma_{x_i}^2 + \sigma_{\hat{x}_i^2}}
         \cdot \Delta x_i \;.
\end{displaymath}
The variance of the residual in this case is therefore
\begin{equation}
\sigma_{\rm incl}^2 \equiv \left<(\Delta x_i^{\prime})^2\right> =
 \left<(x_i - \hat{x}_i^\prime)^2\right> =
        \frac{\sigma_{x_i}^4}{\sigma_{\hat{x}_i}^2 + \sigma_{x_i}^2} \label{a2}\;.
\end{equation}

Combining Eq.\,(\ref{a1}) and Eq.\,(\ref{a2}), one gets the following relation
\begin{displaymath}
\sigma_{x_i}^4 = \left<(\Delta x_i)^2\right>
                \cdot \left<(\Delta x_i^{\prime})^2\right> \;,
\end{displaymath}
which immediately gives the expression for the {\it true} spatial resolution:
\begin{equation}
\sigma_{x_i} = \sqrt{\sigma_{\rm excl} \cdot \sigma_{\rm incl}} \label{a3}\;.
\end{equation}

It should be noted here that Eq.\,(\ref{a3}) is valid not only for straight line
fitting as dealt with in the appendix of Ref.\,\cite{carleton} 
but also for track fitting with any function such as circles used in our analysis.  


\end{document}